\documentclass[letterpaper, 10pt, conference]{ieeeconf}  

\usepackage{graphicx}
\usepackage{amsmath}
\IEEEoverridecommandlockouts                              

\title{\LARGE \bf
Evaluation of Automated Vehicles in the Frontal Cut-in Scenario - an Enhanced Approach using Piecewise Mixture Models
}

\author{Zhiyuan Huang$^{1}$, Ding Zhao$^{2}$, Henry Lam$^{1}$, David J. LeBlanc$^{2}$ and Huei Peng$^{3}$
\thanks{*The first two authors, Z. Huang and D. Zhao, have equally contributed to this research. This work is funded by the Mobility Transformation Center at the University of Michigan with grant No. N021552}
\thanks{$^{1}$Zhiyuan Huang ({\tt\small zhyhuang@umich.edu}) and $^{3}$Henry Lam ({\tt\small khlam@umich.edu}) are with Department of Industrial and Operations Engineering at the University of Michigan, Ann Arbor, MI 48109, USA}%
\thanks{$^{2}$Ding Zhao ({\tt\small zhaoding@umich.edu}) and David LeBlanc ({\tt\small leblanc@umich.edu}) are with the University of Michigan Transportation Research Institute, Ann Arbor, MI 48109}%
\thanks{$^{3}$Huei Peng ({\tt\small hpeng@umich.edu}) is with Department of Mechanical Engineering at the University of Michigan Transportation Research Institute, Ann Arbor, MI 48109}%
}

\begin{document}

\maketitle
\thispagestyle{empty}
\pagestyle{empty}

\begin{abstract}
Evaluation and testing are critical for the development of Automated Vehicles (AVs). Currently,  companies test AVs on public roads, which is very time-consuming and inefficient. We proposed the Accelerated Evaluation concept which uses a modified statistics of the surrounding vehicles and the Importance Sampling theory to reduce the evaluation time by several orders of magnitude, while ensuring the final evaluation results are accurate. In this paper, we further extend this idea by using Piecewise Mixture Distribution models instead of Single Distribution models. We demonstrate this idea to evaluate vehicle safety in lane change scenarios. The behavior of the cut-in vehicles was modeled based on more than 400,000 naturalistic driving lane changes collected by the University of Michigan Safety Pilot Model Deployment Program. Simulation results confirm that the accuracy and efficiency of the Piecewise Mixture Distribution method are better than the single distribution.  \\
%
%
%

\end{abstract}


\section{INTRODUCTION}

Automated and robotics systems requires thoroughly and rigorously evaluation to avoid risks. Recent crashes involving a Google self-driving car \cite{GoogleAutoLLC} and Tesla Autopilot vehicles \cite{PreliminaryHWY16FH018} brought the Public's attention to Automated Vehicle (AV) testing and evaluation. While these AVs are generally considered as among the best current technologies, because they use public road testing, statistically they have not yet accumulated enough miles. The Tesla Autopilot in particular was criticized for being released too early in the hands of the general public \cite{Evan2016FatalPerfect}. 

Currently, there are no standards or protocols to test AVs at automation level 2 \cite{U.S.DepartmentofTransportation2016} higher. Many companies adopt the Naturalistic Field Operational Tests approach \cite{FESTA-Consortium2008}. However, this method is inefficient because safety critical scenarios rarely happen in daily driving. The Google Self-driving cars accumulated 1.9 million driving. This distance, although sounds a lot, provides limited exposure to critical events, given that U.S. drivers encounter a police reported crash every five hundred thousand miles on average and fatal crash every one hundred million miles \cite{NHTSA2014b}. In the meantime, both Google and Tesla update their software throughout the process, which may have improved safety, but the newest version of the AV has not accumulated that many miles as they have claimed.  In summary, today's best practice adopted by the industry is time-consuming and inefficient.  A better approach is needed. 

We proposed the Accelerated Evaluation concept \cite{Zhao2016f} to provide a brand-new alternative. The basic concept is that as high-level AVs just began to penetrate the market, they mainly interact with human-controlled vehicles (HVs). Therefore we focus on modeling the interaction between the AV and the HV around it.  The evaluation procedure involves four steps:
\begin{itemize}
	\item Model the behaviors of the “primary other vehicles” (POVs) represented by Probability Density Function (PDF) $f(x)$ as the major disturbance to the AV using large-scale naturalistic driving data
	\item Skew the disturbance statistics from $f(x)$ to modified statistics $f^{*}(x)$ (accelerated distribution) to generate more frequent and intense interactions between AVs and POVs
	\item Conduct “accelerated tests” with $f^{*}(x)$ 
	\item Use the Importance Sampling (IS) theory to “skew back” the results to understand real-world behavior and safety benefits
\end{itemize}

This approach has been successfully applied to evaluate AVs in the frontal crash with a cut-in vehicle \cite{Zhao2016AcceleratedTechniques} and also frontal crash with a lead vehicle \cite{Zhao2015i,Zhao2016g}. This approach was confirmed to significantly reduce the evaluation time while accurately preserving the statistical behavior of the AV-HV interaction. In the previous studies, the evaluation time was reduced by two to five orders of magnitudes -  the accelerated rate depends on the test scenarios.  In our observation, rarer events achieve higher accelerated rate.  

\begin{figure}[ht]
      \centering
   \includegraphics[width=\linewidth]{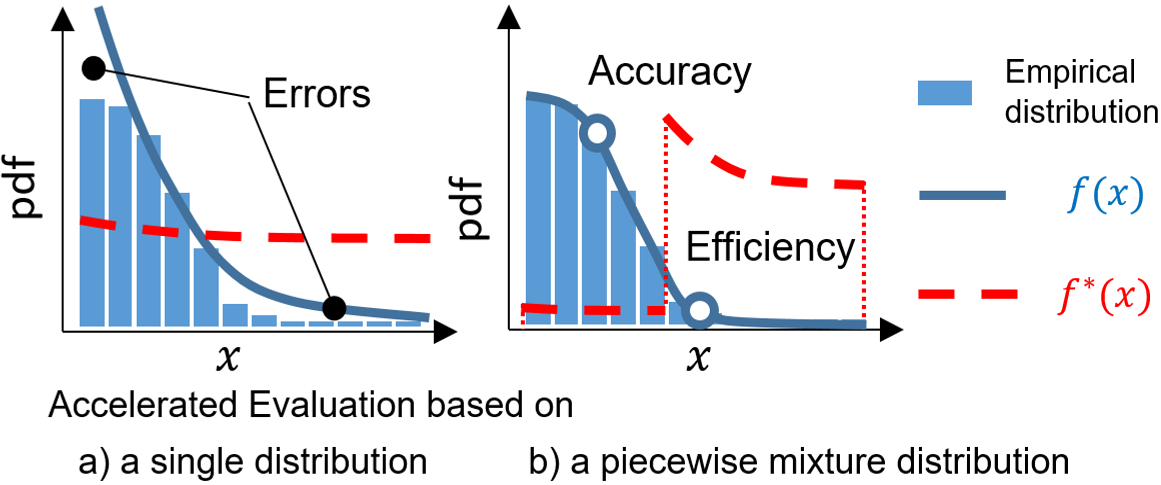}
      \caption{Acceleration evaluation based on single distribution and Piecewise Mixture Distribution.}
      \label{fig:g1}
\end{figure}

Fig. \ref{fig:g1} gives a sketch of the contribution of our paper. Piecewise Mixture Distribution Model is proposed for fitting $f(x)$ and constructing $f^*（x）$ in the above procedure. In our previous studies, the non-accelerated models and the accelerated models were built based on signal component distributions. While this method does benefit from its simple mathematical form, it has a few drawbacks as illustrated in Fig. \ref{fig:g1}. i) The fitting of the rare events (usually the tail part of the statistical distributions) would be dominated by the fitting of the normal driving behaviors (the majority part of the distributions), which may induce large errors. ii) The full potential in higher accelerated rate is not achieved due to the lack of model flexibility. In this paper, we proposed a more general framework for the Accelerated Evaluation method to overcome the aforementioned limitations based on Piecewise Mixture Distribution Models as illustrated in Fig. \ref{fig:g1} b).

We will demonstrate this method using the cut-in scenario. Section \ref{sec:single} will introduce the lane change model based on single distributions. In Section \ref{mixture_section}, we present the new lane change model with Piecewise Mixture Distributions. In Section \ref{sec:accelerated_eval}, the Accelerated Evaluation will be established with simulation results discussed in Section \ref{sec_simulation}. Section \ref{sec:conclusions} concludes this paper.


\section{ACCELERATED EVALUATION WITH SINGLE DISTRIBUTIONS}
\label{sec:single}
Here, we review our previous work in \cite{Zhao2016AcceleratedTechniques}. The lane change events were extracted from the Safety Pilot Model Deployment (SPMD) database \cite{Bezzina2014}.  With over 2 million miles of vehicle driving data collected from 98 cars over 3 years, 403,581 lane change events were identified. We used 173,692 events with negative range rate to build the statistical model focusing on three key variables that captured the effects of gap acceptance of the lane changing vehicle. These three variables are velocity of the lead vehicle ($v_L$), range to the lead vehicle ($R_L$) and time to collision ($TTC_L$). The $TTC_L$ is defined as:
\begin{equation}
	TTC_L=- \frac{R_L}{\dot{R_L}},
\end{equation}
where $\dot{R_L}$ is the relative speed. Since $TTC_L$ is dependent on $v_L$, the data was split into 3 segments: $v_L$ at 5 to 15 m/s, 15 to 25 m/s and 25 to 35 m/s. $R_L$ is independent on $v_L$, and is independent with  $TTC_L$. Therefore the procedure starts with sampling $v_L$ from the empirical distribution, and then we sample from distribution model of $TTC_L$ and $R_L$. By comparing among 17 types of commonly used distribution templates, the Pareto distribution was selected to model $R_L^{-1}$ and the exponential distribution was used for $TTC_L^{-1}$ segments.

\begin{figure}[thpb]
      \centering
   \includegraphics[width=\linewidth]{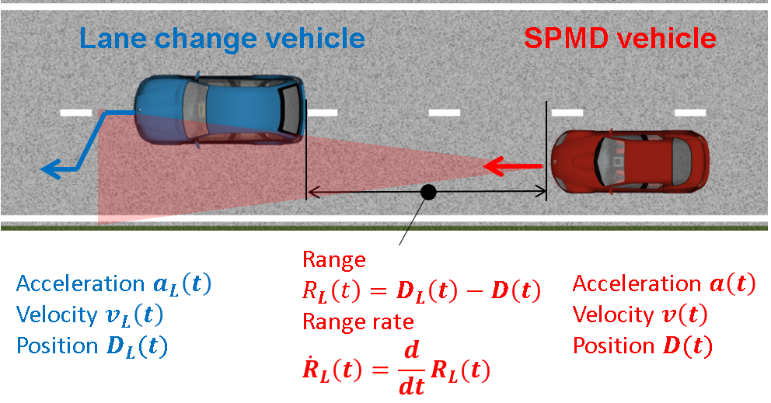}
      \caption{Lane change data collected by SPMD vehicle.}
      \label{fig:o1}
\end{figure}

Given all the key parameter values $v_L$, $R_L$ and $TTC_L$, we input them to the model to simulate the AV-HV interaction. The outcome from the simulation can be considered as an event indicator function $I_\varepsilon(x)$ that returns $\{1, 0\}$ depending on the event of interest. The event indicator function is simulated by AV models using Adaptive Cruise Control (ACC) and Autonomous Emergency Braking (AEB) \cite{Ulsoy2012a} regarding the surrounding environment. Given the stochastic distribution of the variables and the event indicator function, the optimal exponential distribution for Importance Sampling was obtained by implementing the Cross Entropy method \cite{rubinstein1999rare}.

\section{LANE CHANGE MODEL WITH PIECEWISE MIXTURE DISTRIBUTIONS }
\label{mixture_section}

Although many commonly used distributions have concise and elegant forms, they do not always describe the data distribution well. We choose to achieve better fitting the data in subsets. We estimate the model parameters using the Maximum Likelihood Estimation (MLE)\cite{Aldrich1997RA1912-1922}. The process is described below.

Let us assume that we have a family of distribution with Cumulative Distribution Function (CDF) $F(x|\theta)$, where $\theta$ is the parameter vector of $F$. The corresponding Probability Density Function (PDF) of $F$ is $f(x|\theta)$. Assuming that data $D=\{X_1, X_2,...,X_n\}$ is independently and identically distributed and the distribution is in the family of $F(x|\theta)$, we want to find the most ``likely" parameter $\hat{\theta}$.

We define the likelihood function as \begin{equation}
	L(\theta|D)=P(D|\theta)=\Pi_{i=1}^{n} f(X_i|\theta).
\end{equation}
then the estimation of $\hat{\theta}$ that maximizes the likelihood function is called mostly likely estimation MLE.

For computation convenience we introduce the log-likelihood function, which writes as
    \begin{equation}
		l(\theta|D)=\log L(\theta|D)=\sum_{i=1}^{n} \log f(X_i|\theta).
	\end{equation}	
Since $log$ is monotone, the log-likelihood function preserves the optimizer of the original function. The optimizer of log-likelihood function, $\hat{\theta}$, is the MLE of distribution family $F$. We have the MLE as \begin{equation}
\hat{\theta} = \arg \max_{\theta}  \  l(\theta|D).
\end{equation}

In the following, we will describe the idea of  Piecewise Mixture Distribution fitting. Then we present bounded distribution fitting results. The optimization problem is solved by {\bf fminunc} in MATLAB.

	\subsection{General Framework of the Piecewise Mixture Distribution Lane Change Model}
We define Piecewise Mixture Distribution to be distribution with CDF in the form of 
\begin{equation}
\label{eq:CDF}
		F(x)=\sum_{i=1}^{k} \pi_i F_i(x |\gamma_{i-1} \leq x< \gamma_i),
\end{equation}
where $k$ is the number of truncation, $\sum_{i=1}^{k}  \pi_i=1$ and $F_i(x |\gamma_{i-1} \leq x< \gamma_i)$ is conditional cumulative distribution function, which means that $F_i(\gamma_{i-1} |\gamma_{i-1} \leq x< \gamma_i)=0$ and $F_i(\gamma_i |\gamma_{i-1} \leq x< \gamma_i)=1$ for $i=1,...,k$.
	
	We can consider that $\pi_i=P(\gamma_{i-1} \leq x< \gamma_i)$ and when $x \geq 0$, we have $\gamma_0=0$ and $\gamma_k=\infty$.
	
	By this definition, the PDF of Piecewise Mixture Distribution is \begin{equation}
	f(x)=\sum_{i=1}^{k} \pi_i f_i(x |\gamma_{i-1} \leq x< \gamma_i).
\end{equation}

%
%
%
%
%
%
	
	In our case, we have $\theta= \{\pi_1,...,\pi_k, \theta_1,...,\theta_k\}$, where $\theta_i$ is the parameter(s) for $F_i$. By splitting $D$ into pieces regarding the truncation points $\{\gamma_1,...,\gamma_{k-1}\}$, we have data index sets $S_i=\{j|\gamma_{i-1} \leq X_j<\gamma_i\}$ for $i=1,...,k$. We can write the log-likelihood function as
    \begin{equation}
    \label{eq:mle}
    	\begin{array}{l}
		l(\theta|D)  =\sum_{i=1}^{k}     \sum_{j \in S_i} \log \pi_i	\\
      \hspace{3em} +\sum_{i=1}^{k}     \sum_{j \in S_i} \log f_i(X_j  | \gamma_{i-1} \leq x< \gamma_i,\ \theta_i) .
        \end{array}
	\end{equation}
    
	MLE of $\theta$ can be obtained by maximizing $l(\theta|D)$ over $\theta$. Since $l$ is concave over $\pi_i$, we take \begin{equation}
	\frac{\partial l}{\partial \pi_i} =0
\end{equation}
%
%
%

%
%
    and get \begin{equation}
	\hat{\pi_i} = |S_i|/n. 
\end{equation}

	For parameters $\theta_i$ in $F_i$, it was known (\ref{eq:mle}) to be the same as computing MLE of $\theta_i$ with corresponding data set $D_i=\{X| \gamma_{i-1} \leq X< \gamma_i  \ and \ X \in D\}$. Since we use bounded distribution for each $F_i$, we will discuss the estimation of parameters for specific distributions that we applied in later sections.
	
	\subsection{Bounded Distribution}
We will introduce the development of three bounded distributions and use them in the lane change model development.
	
	\subsubsection{MLE for bounded exponential distribution}
	
The bounded exponential distribution has the form\begin{equation}
	f(x|\gamma_1 \leq x < \gamma_2) = \frac{\lambda e ^{-\lambda x}}{e^{-\lambda \gamma_1}-e ^{-\lambda \gamma_2}}
\end{equation}
for $\gamma_1 \leq x < \gamma_2$. 
    
    For data set $D=\{X_1,...,X_n\}$, the log-likelihood function is\begin{equation}
	l(D|\theta)=\sum_{i=1}^{n} \log \lambda -\lambda X_i - \log(e ^{-\lambda \gamma_1}-e ^{-\lambda \gamma_2}),
\end{equation}
where $l$ is concave over $\lambda$. Although we cannot solve the maximization analytically, it is solvable through numerical approach.
	
	Therefore, the MLE of $\lambda$ is \begin{equation}
	\lambda=\arg \max_\lambda  \  n\log \lambda - n \log(e ^{-\lambda \gamma_1}-e ^{-\lambda \gamma_2})-\sum_{i=1}^{n}\lambda X_i.
\end{equation}

	\subsubsection{MLE for bounded normal distribution}
	
	Let us consider that bounded normal distribution with mean 0 and variance $\sigma^2$ conditional on $0 \leq \gamma_1 \leq x < \gamma_2 $. The PDF is \begin{equation}
	f(x|\gamma_1 \leq x < \gamma_2)= \frac{\frac{1}{\sigma}\phi(\frac{x}{\sigma})}{\Phi(\frac{\gamma_2}{\sigma})-\Phi(\frac{\gamma_1}{\sigma})}.
\end{equation}
	
	Objective function for MLE of bounded normal distribution is\begin{equation}
	\max_\sigma -\frac{\sum_{i=1}^{n} X_i^2}{2\sigma^2}-n\log\sigma-n\log(\Phi(\frac{\gamma_2}{\sigma})-\Phi(\frac{\gamma_1}{\sigma})).
\end{equation}

\subsection{Fitting Mixture Model with EM Algorithm}
Comparing to single distributions, mixture distribution can combine several classes of distribution and thus is more flexible. In this section, we discuss the fitting problem of mixture bounded normal distribution.


The PDF of mixture of $m$ bounded normal distribution can be written as\begin{equation}
f(x|\gamma_1 \leq x < \gamma_2)=\sum_{j=1}^{m} p_j f_j(x|\gamma_1 \leq x < \gamma_2)
\end{equation}
where $f_j$ is bounded Gaussian distribution with mean 0 and variance $\sigma_j$. We want to find MLE of $p_j$ and $\sigma_j$ for $j=1,...,m$.

The log-likelihood function for data $D=\{X_i\}_{i=1}^n$ is\begin{equation}
l(\theta|D)=\sum_{i=1}^{n} \log  \sum_{j=1}^{m} p_j f_j(X_i|\gamma_1 \leq x < \gamma_2).
\end{equation}
Since there is a sum within log function, this is hard to optimize. Therefore we need to introduce Expectation-Maximization (EM) algorithm to find the optimizer, i.e. MLE, for the parameters.
%
%
%

We define $Z_i^j$ to denote whether or not the random number $X_i$ comes from mixture distribution $j$, $Z_i^j=\{0, 1\}$. We also introduce the expectation\begin{equation}
E[Z^j_i|X_i]:=\tau_i^j.
\end{equation}

EM \cite{Dempster1977b} algorithm starts with initial parameters $\{p_j,\sigma_j\}$, $j=1,...,m$. For data $D=\{X_i\}_{i=1}^n$, we set complete data as $D_c=\{X_i, Z_i\}_{i=1}^n$. EM algorithm optimizes $E[l_c(\theta|D_c)|D]$ in every step. E step is to update the function $E[l_c(\theta|D_c)|D]$, M step is to optimize this function. The algorithm iterates E step and M step until the convergence criterion is reached.

In our case, we have \begin{equation}
\begin{array}{l}
E[l_c(\theta|D_c)|D]=\sum_{i=1}^{n} \sum_{j=1}^{m} \tau_i^j (\log p_j + \log f_j(X_i)). 
\end{array}
\end{equation}

Objective $E[l_c(\theta|D_c)|D]$ for M step is concave over $p_j$ and $\sigma_j$, we could maximize the objective function through an analytic approach for $p_j$:	\begin{equation}
	p_j=\frac{\sum_{i=1}^{n} \tau_i^j }{n}.
\end{equation}
For $\sigma_j$, we can solve the following maximization problem through numerical approach.
\begin{equation}
\resizebox{\hsize}{!}{$
	\sigma_j=\arg \min_{\sigma_j} \tau_i^j (-\log \sigma_j+ \log \phi(\frac{X_i}{\sigma_j})- \log (\Phi(\frac{\gamma_2}{\sigma_j})-\Phi(\frac{\gamma_1}{\sigma_j}))).$}
\end{equation}

 The full algorithm is presented in the Appendix A.

\section{ACCELERATED EVALUATION WITH IMPORTANCE SAMPLING}
\label{sec:accelerated_eval}

Crude Monte Carlo simulations for rare events can be time consuming. Importance Sampling is thus used to accelerate the evaluation process. The Importance Sampling concept is reviewed below. 

Let $x$ be a random variable generated from distribution $F$, and $\varepsilon \subset \Omega$ where $\epsilon$ is the rare event of interest and $\Omega$ is the sample space. Our objective is to estimate \begin{equation}
{P}(x \in \varepsilon)=E[I_\varepsilon(x)]=\int I_\varepsilon(x) dF
\end{equation}
where\begin{equation}
	I_\varepsilon(x)=\begin{cases} 1 & x \in \varepsilon,\\
0 & otherwise.\end{cases}
\end{equation}

The evaluation of rare events can be written as the sample mean of $I_\varepsilon(x)$: \begin{equation}
	\hat{P}(x \in I) = \frac{1}{N} \sum_{i=1}^N I_\varepsilon(X_i),
\end{equation}
where $X_i$'s are drawn from distribution $F$.

For any distribution $F^*$ that has the same support with $F$, we have \begin{equation}
	E[I_\varepsilon(x)]=\int I_\varepsilon(x) dF = \int I_\varepsilon(x) \frac{dF}{dF^*} dF^* ,
\end{equation}
we can compute the sample mean of $I_\varepsilon(x) \frac{dF}{dF^*}$ over the distribution $F^*$ to have an unbiased estimation of ${P}(x \in I) $. By appropriately selecting $F^*$, the later evaluation procedure provides estimation with smaller variance. This is known as Importance Sampling \cite{Bucklew2004a}.

Exponential change of measure is commonly used to construct $F^*$. Although the exponential change of measure cannot guarantee convergence to optimal distribution, it is easy to implement and the new distribution generally stays within the same class of distribution.

Exponential change of measure distribution is in the form of \begin{equation}
f_\theta(x)=exp( \theta x - \kappa (\theta))f(x),
\end{equation}
where $\theta$ is the change of measure parameter and $\kappa(\theta)$ is the log-moment generating function of original distribution $f$.

For conditional exponential distribution, the exponential change of measure distribution is \begin{equation}
	F_\theta(x| \gamma_1 < x < \gamma_2)= \frac{F_\theta(x)-F_\theta(\gamma_1)}{{F_\theta}(\gamma_2)-{F_\theta}(\gamma_1)}.
\end{equation}

For conditional normal distribution, the exponential change of measure distribution is \begin{equation}
	F_\theta(x| \gamma_1 < x < \gamma_2)=\frac{\Phi(\frac{x-\sigma^2 \theta}{\sigma})-\Phi(\frac{\gamma_1-\theta \sigma^2}{\sigma})}{\Phi(\frac{\gamma_2-\theta \sigma^2}{\sigma})-\Phi(\frac{\gamma_1-\theta \sigma^2}{\sigma})}.
\end{equation}
	
%
%
%
We can use Cross Entropy method to estimate the optimal parameter $\theta^*$ for importance sampling.

%
%
%

\section{SIMULATION ANALYSIS}
\label{sec_simulation}
\subsection{Model of the Automated Vehicle}

Applying the model discussed in Section \ref{mixture_section}, we have different choices of distribution and number of truncation for Piecewise Mixture Models. We first present our Piecewise Mixture Models for $R^{-1}$ and $TTC^{-1}$ and then compare the results with single distribution model used in \cite{Zhao2016AcceleratedTechniques}.
\subsubsection{Piecewise Mixture Models for $R^{-1}$ and $TTC^{-1}$}

\begin{figure}[t]
      \centering
   \includegraphics[width=\linewidth]{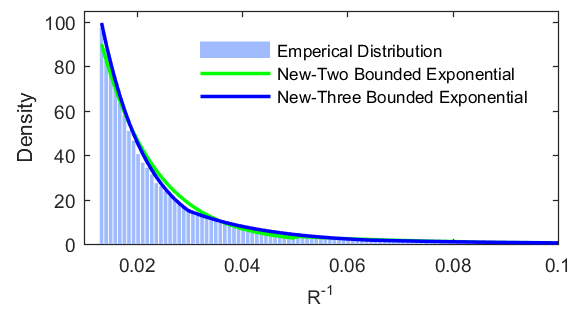}
      \caption{Piecewise Mixture Distribution fitting for $R^{-1}$.}
      \label{fig:r_new}
\end{figure}

The dependence to $R^{-1}$ looks very close to an exponential distribution. The fitting using two bounded exponential distributions and three bounded exponential distributions are presented in Fig. \ref{fig:r_new}. Since truncated one more part would not increase much complexity while providing a more accurate fitting, we choose to use three bounded exponential distributions to model $R^{-1}$.

\begin{figure}[b]
      \centering
   \includegraphics[width=\linewidth]{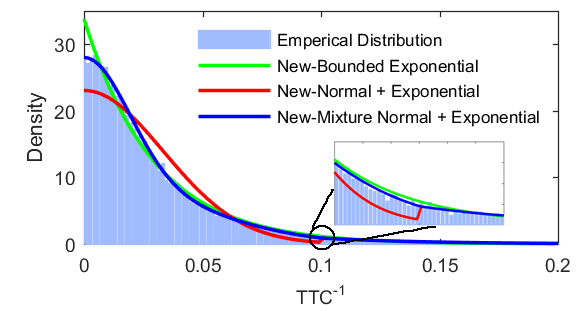}
      \caption{Piecewise Mixture Distribution fitting for $TTC^{-1}$ given $v_L$ between 15 and 25 m/s.}
      \label{fig:ttc2_new}
\end{figure}

The three data segments of $TTC^{-1}$ have similar distributions, therefore we only show our model of the segment for $v_L$ in the range of 15 to 25 m/s. In Fig. \ref{fig:ttc2_new}, we truncated the data into two parts. For the tail part, we use the exponential distribution. For the body part, the mixture of two normal distributions gives a better fit. We also notice that the fitting of piecewise distribution is not continuous at the truncation point.


\subsubsection{Comparison with Single Distribution Models}

\begin{figure}[thpb]
      \centering
   \includegraphics[width=\linewidth]{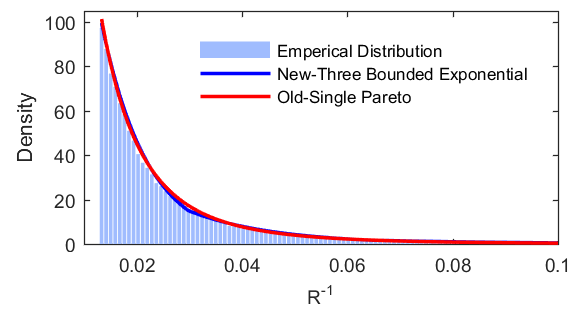}
		\caption{Comparison of fitting for $R^{-1}$.}
      \label{fig:r_old}
\end{figure}

\begin{figure}[thpb]
      \centering
   \includegraphics[width=\linewidth]{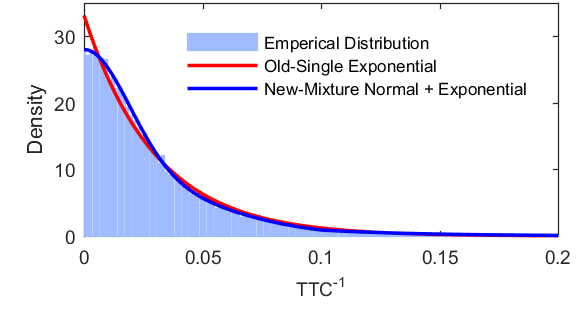}
      \caption{Comparison of fitting for $TTC^{-1}$ given $v_L$ between 15 and 25 m/s.}
      \label{fig:ttc2_old}
\end{figure}

The comparison of new model and the previous model are shown in Fig.\ref{fig:r_old} and Fig.\ref{fig:ttc2_old}. Piecewise Mixture Models provided more flexibility in data fitting, while not bringing much computation burden. 
%
%
%
%
%


\subsection{Simulation Results}

In simulation experiments, we set the convergence criterion for the relative half-width of $100(1-\alpha)\%$ confidence interval, i.e., the simulation terminates when the relative half-width of $100(1-\alpha)\%$ confidence interval gets below $\beta$. In this paper, we use $\alpha=0.2$ and $\beta=0.2$. Based on this convergence criterion, we can study the number of samples needed for convergence. We compare the efficiency of the Piecewise Mixture Distribution and single exponential distribution.

\begin{figure}[b]
      \centering
   \includegraphics[width=\linewidth]{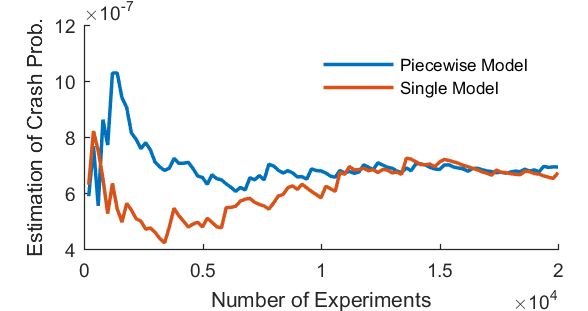}
      \caption{Estimation of crash probability for one lane change using piecewise and single accelerated distributions.}
      \label{fig:result_estimation}
\end{figure}

\begin{figure}[thpb]
      \centering
   \includegraphics[width=\linewidth]{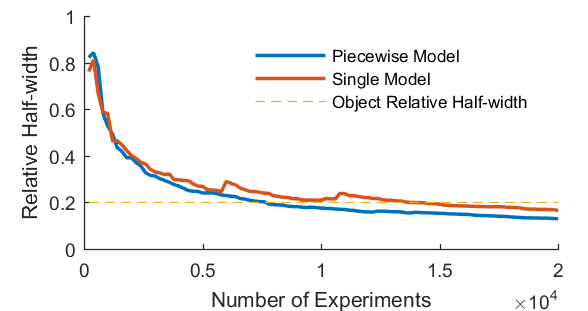}
      \caption{The relative half-width for simulation with piecewise and single accelerated distributions.}
      \label{fig:result_relative_hw}
\end{figure}
%
%
Fig. \ref{fig:result_estimation} shows that both models give a similar estimation as the number of experiments gets large and the Piecewise Mixture Distribution model converges slightly faster than the single model. Fig. \ref{fig:result_relative_hw} shows that the relative half-width of the Piecewise Mixture Distribution model reaches the target confidence value after 7800 samples and the single model needs about 13800 samples, a reduction of 44\%.

When we repeat the process for 10 times, it takes on average 7840 samples to get a converged estimation using piecewise accelerated distribution, while the single accelerated distribution needs on average 12320 samples to converge. Table \ref{table:n_ratio} presents the comparison of these two approaches with the Crude Monte Carlo method\cite{Asmussen2007StochasticAnalysis}. The number needed for convergence of crude Monte Carlo is computed using the fact that the number of occurred event of interest is Binomial distributed. We can compute the standard deviation of the crude Monte Carlo estimation $\hat{P}(x \in \varepsilon)$ by \begin{equation}
std(\hat{P}(x \in \varepsilon))=\sqrt{\frac{\hat{P}(x \in \varepsilon)(1-\hat{P}(x \in \varepsilon))}{n}}.
\end{equation} 
Then we can estimate \begin{equation}
\hat{N}=\frac{ z_{\alpha/2}^2 (1-\hat{P}(x \in \varepsilon))}{ \beta^2  \hat{P}(x \in \varepsilon)}, 
\end{equation}
where $z_{\alpha/2}$ is the $(1-\alpha/2)$ quantile of normal distribution. The required sample size $N$ of crude Monte Carlo in Table \ref{table:n_ratio} is calculated from an estimation $\hat{P}(x \in \varepsilon)=7.4\times 10^{-7}$ with $80\%$ confidence interval $(7.0\times 10^{-7},7.8\times 10^{-7})$.
\begin{table}[t]
\centering
\caption{The table shows the number of samples (N) needed to reach the convergence criterion and the ratio normalized against the piecewise accelerated distribution. }
\label{table:n_ratio}
\begin{tabular}{|l|l|l|l|}
\hline
           & Piecewise   & Single & Crude     \\ \hline
N           & 7840 & 12320  & $5.5\times 10^7$ \\ \hline
Ratio to Mixture & 1    & 1.57   & $7\times 10^3$   \\ \hline
\end{tabular}
\end{table}

\section{CONCLUSIONS}
\label{sec:conclusions}
This study proposes a new model for accelerated evaluation of AVs. The Piecewise Mixture Distribution Models provide more accurate fitting to the vehicle behaviors than the single model used in the literature, and was found to be more efficient. In rear-end crashes caused by improper lane changes studied in this article, the Piecewise Mixture Distribution model reduces the simulation cases by about 33\% compared with the single model under the same convergence requirement, and is 7000 times faster than the Crude Monte Carlo method. 
%
%
%
%
\addtolength{\textheight}{-1cm}   



\section*{APPENDIX}

\subsection{EM Algorithm for Mixture Bounded Normal Distribution}

Here we present a numerical MLE algorithm with mixture bounded normal distribution.

{\bf ALGORITHM:}
%
%
%
\begin{enumerate}
	\item Initial $\{p_j,\sigma_j\}$, $j=1,...,m.$
	\item E step: update \begin{equation}
\tau_i^j= \frac{p_j f_j(X_i|\sigma_j)}{\sum_{j=1}^{m}p_j f_j(X_i|\sigma_j)}.
\end{equation} 
	\item M step: update \begin{equation}
	p_j=\frac{\sum_{i=1}^{n} \tau_i^j }{n}
\end{equation}
    and 
	\begin{equation}
	\sigma_j=\arg \min_{\sigma_j} - \tau_i^j \log \frac{\sigma_j (\Phi(\frac{\gamma_2}{\sigma_j})-\Phi(\frac{\gamma_1}{\sigma_j}))}{\phi(\frac{X_i}{\sigma_j})}.
\end{equation}

	\item Repeat 2 and 3 until $l(\theta|D)$ converges.
\end{enumerate}

\subsection{Inverse CDF of Piecewise Mixture Distributions}
	
	We could sample from Piecewise Mixture Distribution by the inverse CDF approach. Here we present the inverse CDF for Piecewise Mixture Distribution. 
	
	The CDF of Piecewise Mixture Distribution (\ref{eq:CDF}) can split into\begin{equation}
    \resizebox{\hsize}{!}{$
F(x)= \begin{cases} 
	... \\
	\sum_{j=1}^{i-1}\pi_j + \pi_i F_i(x | \gamma_{i-1} \leq x< \gamma_i) &  \gamma_{i-1} \leq x< \gamma_i\\
	...
	\end{cases}.$}
\end{equation}
Therefore the inverse function can be written as\begin{equation}
\label{eq:icdf}
\resizebox{\hsize}{!}{$F^{-1}(y)=\begin{cases} 
	... \\
	F_i^{-1}(\frac{y-\sum_{j=1}^{i-1}\pi_j}{\pi_i} | \gamma_{i-1} \leq x< \gamma_i) & \sum_{j=1}^{i-1}\pi_j \leq y < \sum_{j=1}^{i}\pi_j\\
	...
	\end{cases}.$   }
\end{equation}
where $F_i^{-1}$ is the inverse conditional CDF of $F_i$. We present two example of inverse conditional CDF in the following.
	
For inverse CDF of conditional exponential distribution, we have
\begin{equation}
\begin{array}{ll}
F_\theta^{-1}(y|F_\theta(\gamma_1)\leq y<F_\theta(\gamma_2))\\=F_\theta^{-1}((F_\theta(\gamma_2)-F_\theta(\gamma_1))y+F_\theta(\gamma_1)),
\end{array}
\end{equation}
where $F$ and $F^{-1}$ are CDF and inverse CDF of exponential distribution.

	For conditional normal distribution, the inverse CDF is\begin{equation}
	\begin{array}{ll}
&F_\theta^{-1}(y|F_\theta(\gamma_1) \leq y < F_\theta(\gamma_2))   \\ 
=&\sigma \Phi^{-1} ((\Phi(\frac{\gamma_2-\theta \sigma^2}{\sigma})-\Phi(\frac{\gamma_1-\theta \sigma^2}{\sigma}))y \\
&+\Phi(\frac{\gamma_1-\theta \sigma^2}{\sigma}))+\theta \sigma^2.
	\end{array}
\end{equation}
\\



\bibliographystyle{IEEEtran}
\bibliography{root}

\end{document}